\documentclass[preprint,aps,12pt,showpacs,nofootinbib,tightenlines]{revtex4}
\usepackage{amsmath}
\usepackage{amssymb}
\usepackage{epsfig}
\usepackage{graphicx}
\begin{document}
%%%%%%%%%%%%%%%%%%%%%%%%%%%%%%%%%%%%%%%%%%%%%
%=====================================
%              definitions
%=====================================
\def\be{\begin{eqnarray}}
\def\en{\end{eqnarray}}
\def\non{\nonumber\\}

\def\ra{\rangle}
\def\la{\langle}
\def\sl{\!\!\!\slash}

\newcommand{\acp}{{\cal A}_{CP}}
\newcommand{\etap}{\eta^{\prime} }
\newcommand{\etapp}{\eta^{(\prime)}}
%%--------------------------------------------
\title{$B \to f_0(980) (\pi,\etapp)$ Decays in the PQCD Approach }
\author{Zhang Zhi-Qing  and Xiao Zhen-Jun
\footnote{Electronic address: xiaozhenjun@njnu.edu.cn} } %%
\affiliation{Department of Physics and Institute of
Theoretical Physics,
Nanjing Normal University, Nanjing, Jiangsu 210097, P.R.China }
\date{\today}
\begin{abstract}
Based on the assumption of two-quark structure of the scalar meson
$f_0(980)$, we calculate the branching ratios and CP-violating
asymmetries for the four $B \to f_0(980)\pi$ and $B \to
f_0(980)\etapp$ decays by employing the perturbative QCD (pQCD)
factorization approach. The leading order pQCD predictions for
branching ratios are,
$Br(B^-\to f_0(980) \pi^-)\sim 2.5 \times 10^{-6}$,
$Br(\bar{B}^0\to f_0(980) \pi^0)\sim 2 6 \times 10^{-7}$,
$Br(\bar{B}^0\to f_0(980) \eta )\sim 2.5 \times 10^{-7}$ and
$Br(\bar{B}^0\to f_0(980) \eta^\prime)\sim 6.7 \times 10^{-7}$,
which are consistent with both the QCD factorization predictions and the
experimental upper limits.
\end{abstract}


% Key Words: B meson decay; scalar meson; pQCD factorization approach
\vspace{1cm}

\maketitle

%=======================================================================
%                     Introduction
%=======================================================================

Very recently, some $B\to S P$ decays have been studied, for example,
by employing the QCD factorization  (QCDF) approach or the perturbative QCD (PQCD)
approach \cite{cyf0k,chenf0k1,wwang}.
In B factory, the first scalar meson $f_0(980)$ observed
in the decay mode $B\to f_0(980)K$ by Belle \cite{bellef0}, and
confirmed by BaBar \cite{babarf0} later, then many $B\to S P$
channels have been measured \cite{belle1,babar3}.

In this paper, we will calculate the branching ratios and
CP asymmetries of $B^{-} \to f_0(980)\pi^{-}$,$\bar{B}^0\to
f_0(980)\pi^0$ and $\bar{B}^0\to f_0(980)\etapp$ decays in
the pQCD approach at leading order.
This paper is organized as follows: In Sect.~1, we give a brief discussion about
the physical properties of $f_0(980)$, and will calculate the decay amplitudes
for the considered decays. Sect.2 contains the numerical results and discussions.

\section*{1. Decay amplitudes of $B \to f_0(980) (\pi, \etapp)$ decays}
\label{proper}

At present we still do not have a clear understanding about the inner structure
of the scalar mesons.
There are many interpretations for the scalar mesons, such as $qq\bar q \bar q$
four-quark state\cite{jaffe} or $q\bar{q}$ state\cite{nato}, the possibilities of
 $K\bar{K}$ molecular state\cite{jwei}, and even the admixture with glueball states.

In the four-quark model, the flavor wave function of $f_0{(980)}$
 is symbolically given by\cite{jaffe} $f_0=s\bar{s}(u\bar{u}+d\bar{d})/\sqrt{2}$,
 which is supported by a lattice
 calculation. This scenario can explain some experiment phenomena, such as the mass degeneracy of $f_0(980)$  and
 $a_0(980)$, the large coupling of $f_0(980)$ and $a_0(980)$ to
 $K\bar{K}$. But we may wonder if the energetic $f_0(980)$ produced
 in B decays is dominated by the four-quark configuration as it
 requires to pick up two energetic quark-anti quark pairs to form a
 fast-moving light four-quark scalar meson\cite{hycheng1}.

 In the naive 2-quark model, $f_0{(980)}$ is purely an $s\bar{s}$ state
 and this is supported by the data of $D^+_s\to f_0\pi^+$ and $\phi \to f_0\gamma$.
 However, there also exist
 some experiment evidences, such as $\Gamma(J/\psi\to f_0\omega)\approx \frac{1}{2}
 \Gamma(J/\psi\to f_0\phi)$, $f_0(980)\to\pi\pi$ is not OZI suppressed relative to
 $a_0(980)\to\pi\eta$, indicating that $f_0{(980)}$ is not
 purely an $s\bar{s}$ state, but a mixture of
 $s\bar{s}$ and $n\bar{n}\equiv(u\bar{u}+d\bar{d})/\sqrt{2}$:
\be
|f_0(980)\ra = |s\bar s\ra\cos\theta+|n\bar n\ra\sin\theta,
\en
where $\theta$ is the mixing angle. According to
Ref.\cite{hycheng2}, $\theta$ lies in the ranges of $25^\circ <
\theta <40^\circ$ or $140^\circ < \theta < 165^\circ$. Because of our poor knowledge
about the non-perturbative dynamics of QCD, we still can not distinguish between the
four-quark and two-quark model assignment at present.
Some authors, on the other hand,  have shown that the scalar mesons with masses
above 1 GeV can be identified as  conventional $q\bar{q}$ states
with the large possibility\cite{duds,lucd}, this conclusion was obtained by
calculating the masses and the decay constants of these scalar mesons
composed of quark-antiquark pairs based on QCD sum rule.
we here work in the two-quark model and identifying
$f_0(980)$ as the mixture of $s\bar{s}$ and $n\bar{n}$, in order to
give quantitative predictions.

In the two-quark model, the decay constants for scalar meson $f_0(980)$
are defined by:
\be
\langle f_0(p)|\bar q_2\gamma_\mu q_1|0 \ra
&=&0, \quad \langle f_0(p)|\bar q_2 q_1|0\ra=m_S \bar {f_S}.
\en
and
\be
\langle f_0^n|\bar dd|0\ra=\langle f_0^n|\bar uu|0\ra
=\frac{1}{\sqrt 2}m_{f_0}\tilde f^n_{f_0},\quad \langle f_0^s|\bar
ss|0\ra=m_{f_0}\tilde f^s_{f_0},
\en
where $f_0^n$ and $f_0^s$ represent the quark flavor states of $f_0(980)$.
Using the QCD sum rules
method, one can find the scale-dependent scalar decay constants
$f_{f_0}^n$ and $f_{f_0}^s$ are very close\cite{cyf0k,hycheng1}. So
one usually assumes $\tilde f_{f_0}^n=\tilde f_{f_0}^s$ and denotes
them as $\bar f_{f_0}$ in the following.

The twist-2 and twist-3 light-cone distribution amplitudes (LCDAs) for different
components of scalar meson $f_0(980)$ are defined by:
\be
\langle
f_0(p)|\bar q(z)_l q(0)_j|0\rangle
&=&\frac{1}{\sqrt{2N_c}}\int^1_0dxe^{ixp\cdot z}\non && \cdot
\left\{ p\sl\Phi_{f_0}(x)
+m_{f_0}\Phi^S_{f_0}(x)+m_{f_0}(n\sl_+n\sl_--1)\Phi^{T}_{f_0}(x)\right
\}_{jl}. \label{LCDA}
\en
Here we assume that $f_0^n(p)$ and
$f_0^s(p)$ have the same form and denoted as $f_0(p)$, and
$n_+=(1,0,0_T)$ and $n_-=(0,1,0_T)$ are the light-like vectors.

The twist-2 LCDA $\Phi_f(x,\mu)$ can be expanded as the Gegenbauer polynomials:
\be
\Phi_f(x,\mu)&=&\frac{1}{2\sqrt{2N_c}}\bar
f_f(\mu)6x(1-x)\sum_{m=1}^\infty B_m(\mu)C^{3/2}_m(2x-1),
\en
where the values for Gegenbauer moments are taken at scale $\mu=1 \mbox{GeV}$:
$B_1=-0.78\pm0.08$, $B_2=0$ and $B_3=0.02\pm0.07$.

As for the twist-3 distribution amplitudes $\Phi_f^s$ and $\Phi_f^T$,
we adopt the asymptotic form:
\be
\Phi^S_f&=& \frac{1}{2\sqrt {2N_c}}\bar f_f,\,\,\,\,\,\,\,\Phi_f^T=
\frac{1}{2\sqrt {2N_c}}\bar f_f(1-2x).
\en

The B meson is treated as a heavy-light system. We here use the same B meson wave
function as in Ref.~\cite{liu06,guodq07}.
For the $\eta-\etap$ system, we use the quark-flavor basis with
$\eta_q = (u\bar u +d\bar d)/\sqrt{2} $ and $\eta_s=s\bar s$, employ
the same wave function, the identical distribution amplitudes
$\phi_{\eta_{q,s}}^{A,P,T}$, and use the same values for other relevant input
parameters, such as
$f_q=(1.07\pm0.02)f_{\pi}$, $f_s=(1.34\pm0.06)f_{\pi}$, $\phi=39.3^\circ \pm 1.0^\circ, etc$
, as given in Ref.~\cite{feldmann}.
From those currently known studies\cite{liu06,guodq07,ligluon}  we believe that
there is no large room left for the contribution due to the gluonic
component of $\etapp$, and therefore neglect the possible
gluonic component in both $\eta$ and $\etap$ meson.

%===========================================================================
%                    Decay amplitudes in PQCD approach
%============================================================================

The pQCD factorization approach has been used  to study the $ B\to
f_0(980) K $ decays \cite{chenf0k1,wwang}. Following the same
procedure of Ref.~\cite{wwang}, we here would like to study $B \to
f_0(980)\pi $ and $f_0(980)\etapp$ decays by employing the pQCD
approach at leading order.

Since the b quark is rather heavy we consider the $B$ meson at rest
for simplicity. By using the light-cone coordinates the $B$ meson
and the two final state meson's momenta can be written as
\be
P_B =
\frac{M_B}{\sqrt{2}} (1,1,{\bf 0}_T), \quad P_{2} =
\frac{M_B}{\sqrt{2}}(1,0,{\bf 0}_T), \quad P_{3} =
\frac{M_B}{\sqrt{2}} (0,1,{\bf 0}_T),
\en
where the meson masses
have been neglected. Putting the anti- quark momenta in $B$, $P$ and
$S$ mesons as $k_1$, $k_2$, and $k_3$, respectively, we can choose
\be
k_1 = (x_1 P_1^+,0,{\bf k}_{1T}), \quad k_2 = (x_2 P_2^+,0,{\bf
k}_{2T}), \quad k_3 = (0, x_3 P_3^-,{\bf k}_{3T}).
\en

In the pQCD approach, the decay amplitude ${\cal A}(B \to P f_0)$
can be written conceptually as
\be
{\cal A}(B \to P f_0)&\sim& \int\!\! d^4k_1 d^4k_2
d^4k_3\ \mathrm{Tr} \left [ C(t) \Phi_B(k_1) \Phi_{P}(k_2)
\Phi_{f_0}(k_3) H(k_1,k_2,k_3, t) \right ], \non
&\sim &\int\!\!
d x_1 d x_2 d x_3 b_1 d b_1 b_2 d b_2 b_3 d b_3 \non && \cdot
\mathrm{Tr} \left [ C(t) \Phi_B(x_1,b_1) \Phi_{P}(x_2,b_2)
\Phi_{f_0}(x_3, b_3) H(x_i, b_i, t) S_t(x_i)\, e^{-S(t)} \right ],
\quad \label{eq:a2}
\en
where the term ``$\mathrm{Tr}$" denotes the
trace over Dirac and color indices. $C(t)$ is the Wilson
coefficient. The function $H(x_i,b_i,t)$ is the hard part and can be
calculated perturbatively, while $b_i$ is the conjugate space
coordinate of $k_{iT}$, and $t$ is the largest energy scale in hard
function. The function $\Phi_M$ is the wave function which describes
hadronization of the quark and anti-quark to the meson $M$. The
threshold function $S_t(x_i)$ smears the end-point singularities on
$x_i$. The last term, $e^{-S(t)}$, is the Sudakov form factor which
suppresses the soft dynamics effectively.

For our considered decays, the relevant weak effective Hamiltonian
$H_{eff}$ can be written as
\be \label{eq:heff}
{\cal H}_{eff} = \frac{G_{F}} {\sqrt{2}} \, \sum_{q=u,c}V_{qb} V_{qd}^*\left\{
\left [ C_1(\mu) O_1^q(\mu) + C_2(\mu) O_2^q(\mu) \right ]
+ \sum_{i=3}^{10} C_{i}(\mu) \,O_i(\mu) \right\} \; ,
\en
where the Fermi constant
$G_{F}=1.166 39\times 10^{-5} GeV^{-2}$, $V_{ij}$ is the Cabbibo-Kobayashi-Maskawa (CKM)
matrix elements, $C_i(\mu)$ are Wilson coefficients at the renormalization
scale $\mu$ and $O_i$ are the four-fermion operators for the case of
$b \to d $ transition.

In the pQCD approach, the typical Feynman diagrams contributing to
the $\bar{B}^0 \to f_0(980) \pi^0$, $B^- \to f_0(980)\pi^-$ and
$\bar{B}^0\to f_0(980) \etapp$
decays at leading order are illustrated in Fig.~1.
By analytical calculations of the relevant Feynman
diagrams, one can find the total decay amplitudes for the considered decays:
\be
{\cal M}(f_0\;\pi^0) &=&
\frac{\xi_u}{\sqrt{2}}\left[(-M_{e\pi}+M_{a\pi}+M_{ef}+M_{af})C_2+(F_{a\pi}+F_{ef}+F_{af})a_2\right]F_1(\theta)
\non &&
+\frac{\xi_t}{\sqrt{2}} \left\{\left[
F_{e\pi}^{P2}\left (a_6-\frac{1}{2}a_8 \right )
+M_{e\pi}\left (C_3+2C_4-\frac{1}{2}C_9+\frac{1}{2}C_{10}\right )
\right.\right.\non
&&\left.\left.
+ M_{e\pi}^{P2} \left (2C_6+\frac{1}{2}C_8 \right )
+\left (M_{e\pi}^{P1}+M_{a\pi}^{P1}+M_{ef}^{P1}+M_{af}^{P1}\right )\left
(C_5-\frac{1}{2}C_7\right )
\right.\right.\non
&&\left.\left.
+ \left(M_{a\pi}+M_{ef}+M_{af}\right)
\left (C_3-\frac{3}{2}a_{10} \right )
-\left( M_{a\pi}^{P2}+M_{ef}^{P2} +M_{af}^{P2}\right ) \frac{3}{2}C_8
\right.\right.\non &&
\left.\left.
-\left (F_{a\pi}+F_{ef}+F_{af} \right )
\left (-a_4-\frac{3}{2}a_7+\frac{3}{2}a_9+\frac{1}{2}a_{10} \right )
\right.\right.\non
&&
\left.\left.
+\left(F_{a\pi}^{P2}+F_{ef}^{P2}+F_{af}^{P2}\right)
\left (a_6-\frac{1}{2}a_8 \right )\right]F_1(\theta)
\right.\nonumber\\
&&\left.
+\left[M_{e\pi} \left (C_4-\frac{1}{2}C_{10} \right )
+M_{e\pi}^{P2}\left (C_6-\frac{1}{2}C_8 \right )\right]F_2(\theta)\right\}
, \label{eq:m1}
\en
%%%%%%%%%%
\be
{\cal M} (f_0\;\pi^-) &=&
\xi_u\left[M_{e\pi}C_2
+\left (M_{a\pi}+M_{ef}+M_{af} \right )C_1
+\left (F_{a\pi}+F_{ef}+F_{af} \right )a_1\right]F_1(\theta)\non
&&
-\xi_t \left\{\left[
F_{e\pi}^{P2}\left (a_6-\frac{1}{2}a_8 \right )
+ M_{e\pi}\left (C_3+2C_4-\frac{1}{2}C_9+\frac{1}{2}C_{10} \right )
\right.\right.\nonumber\\
&&\left.\left.
+ M_{e\pi}^{P1} \left (C_5-\frac{1}{2}C_7 \right )
+ \left ( M_{a\pi}^{P1}+M_{ef}^{P1}+M_{af}^{P1} \right )
\left ( C_5+C_7 \right )
\right.\right.\nonumber\\
&&\left.\left.
- \left (M_{a\pi}+M_{ef}+M_{af} \right )
\left (C_3+C_9 \right )\right.\right.\non
&&\left.\left.
+ \left (F_{a\pi}+F_{ef}+F_{af} \right )\left (a_4+a_{10} \right )
+ \left (F_{a\pi}^{P2}+F_{ef}^{P2}+F_{af}^{P2} \right )
\left (a_6-\frac{1}{2}a_8 \right )\right]F_1(\theta)
\right.\nonumber\\&&\left.
+ \left [ M_{e\pi}\left (C_4-\frac{1}{2}C_{10} \right )
+ M_{e\pi}^{P2}\left (C_6-\frac{1}{2}C_8 \right ) \right ] F_2(\theta)\right\}
, \label{eq:m2}
\en
%%%%%%%%%%%%%%%
\be
{\cal M} (f_0\;\eta) &=&
\xi_u\left\{\left[(M_{e\eta}+M_{a\eta}+M_{ef}+M_{af})C_2+(F_{a\eta}+F_{af})a_2\right]
+ F_{ef}a_2f_q\right\}F_1(\theta)F_1(\phi)\nonumber\\&&-\xi_t
\left\{\left[ F_{e\eta}^{P2}\left (a_6-\frac{1}{2}a_8 \right )
\right.\right.\non
&&\left.\left.
+(M_{e\eta}+M_{a\eta}+M_{ef}+M_{af})
\left (C_3+2C_4-\frac{1}{2}C_9+\frac{1}{2}C_{10} \right )
\right.\right.\non
&&\left.\left.
+\left (M_{e\eta}^{P1}+M_{a\eta}^{P1}+M_{ef}^{P1}+M_{af}^{P1}\right )
\left (C_5-\frac{1}{2}C_7 \right )
\right.\right.\non
&&\left.\left.
+\left ( M_{e\eta}^{P2}+M_{a\eta}^{P2}+M_{ef}^{P2}+M_{af}^{P2} \right )
\left (2C_6+\frac{1}{2}C_8 \right )
\right.\right.\non
&&\left.\left.
+(F_{a\eta}+F_{ef}f_q+F_{af})
\left (2a_3+a_4-2a_5-\frac{1}{2}a_7+\frac{1}{2}a_9-\frac{1}{2}a_{10} \right )
\right.\right.\nonumber\\&&\left.\left.
+\left (F_{a\eta}^{P2}+F_{ef}^{P2}+F_{af}^{P2} \right )
\left (a_6-\frac{1}{2}a_8 \right )\right]F_1(\theta)F_1(\phi)
\right.\non
&&\left.
+\left [ \left ( F_{a\eta}+F_{ef}f_s+F_{af} \right )
\left (a_3-a_5+\frac{1}{2}a_7-\frac{1}{2}a_9 \right )
\right.\right.\non
&&\left.\left.
+\left (M_{e\eta}+M_{a\eta}+M_{ef}+M_{af} \right )
\left (C_4-\frac{1}{2}C_{10} \right )
\right.\right.\non
&&\left. \left.
+ \left ( M_{e\eta}^{P2} M_{a\eta}^{P2}+M_{ef}^{P2}
+M_{af}^{P2}\right )\left (C_6-\frac{1}{2}C_{8} \right )\right ]
F_2(\theta)F_2(\phi)\right\}
, \label{eq:m3}
\en
where $\xi_u = V_{ub}^*V_{ud}$, $\xi_t = V_{tb}^*V_{td}$, $F_1(\theta)=\sin\theta/\sqrt{2}$ and
$F_2(\theta)=\cos\theta$ are the mixing factors for $f_0(980)$ meson, while
$F_1(\phi)=\cos\phi/\sqrt{2}$ and $F_2(\phi)=-\sin\phi$ are the mixing factors
for $\eta-\etap$ system.
For $B \to f_0(980) \etap$ decay,
the corresponding decay amplitude ${\cal M} (\bar{B}^0 \to f_0\;\etap)$ can be obtained
from  ${\cal M} (\bar{B}^0 \to f_0\;\eta)$ in Eq.~(\ref{eq:m3}) by replacements
of $F_1(\phi) \to F_1^\prime=\sin\phi/\sqrt{2}$
and  $F_2(\phi) \to F_2^\prime=\cos\phi$.

The Wilson coefficients $a_i$ in Eq.~(\ref{eq:m1}-\ref{eq:m3}) are
the combinations of the ordinary Wilson coefficients $C_i(\mu)$,
\be
a_1&=& C_2 + \frac{C_1}{3}, \quad a_2= C_1 + \frac{C_2}{3}, \non
a_{i}&=& C_i + \frac{C_{i+1}}{3}, \ \ {\rm for} \ \ i=3,5,7,9,\non
a_{i}&=& C_i + \frac{C_{i-1}}{3}, \ \ {\rm for} \ \ i=4,6,8,10.
\en

\begin{figure}[t,b]
\vspace{-4cm}
\centerline{\epsfxsize=20cm \epsffile{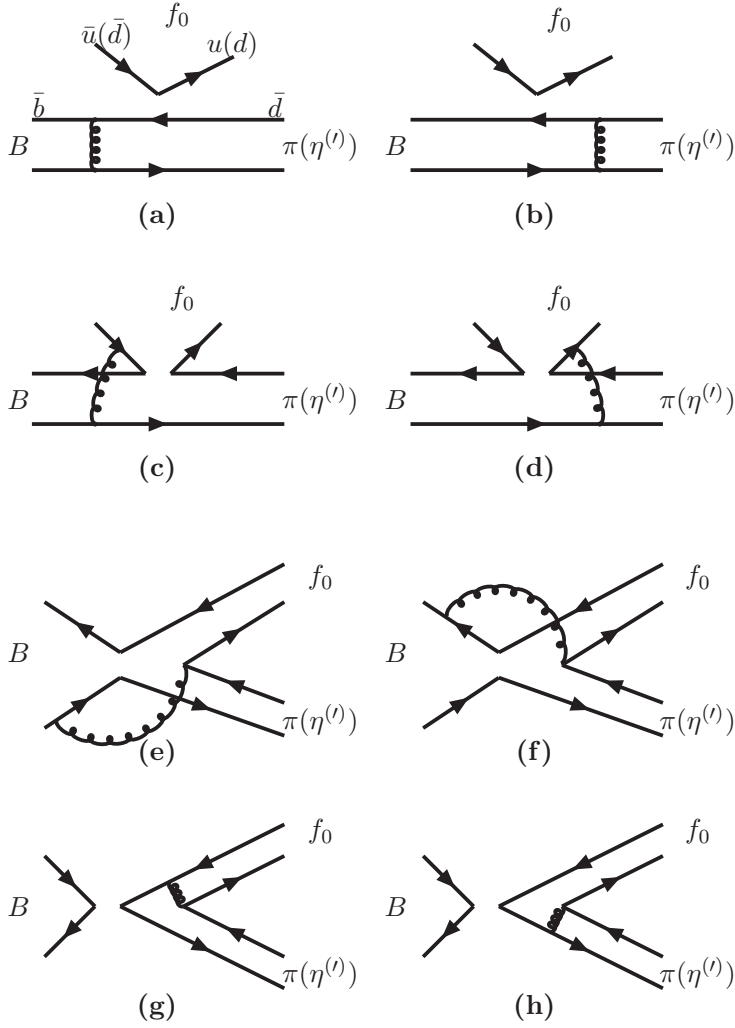}}
\vspace{-10.5cm}
\caption{ Typical Feynman diagrams contributing to the
$B\to  f_0(980) \pi({\etapp})$ decays at leading order .}
\label{fig:fig1}
\end{figure}

The non-zero individual decay amplitudes in Eqs.~(\ref{eq:m1}-\ref{eq:m3}), such as
$F_{e\pi}^{P2}, M_{e\pi}, M_{e\pi}^{P1}, M_{e\pi}^{P2}, \cdots$, are obtained by evaluating
analytically the different Feynman diagrams in Fig.~\ref{fig:fig1}. For
$\bar{B}^0 \to f_0(980)\pi^0$ and $B^- \to f_0(980)\pi^-$ decays, we have
\be
F^{P2}_{e\pi} &=& -16 \pi C_F m_B^4 r_{f}
{\bar{f_{f}}}\int_0^1 dx_1 dx_3 \int_0^{\infty} b_1db_1\, b_3db_3\,
\Phi_B(x_1,b_1)
\nonumber \\
& &\cdot \bigg\{ \left[ \Phi_{\pi}^A(x_3)+r_{\pi} x_3 \left(
\Phi_{\pi}^P(x_3) - \Phi_{\pi}^T(x_3) \right) +2 r_{\pi}
\Phi_{\pi}^P(x_3)
 \right]\nonumber\\
&&\;\;\;\;  \cdot E_{ei}(t) h_{e}(x_1,x_3,b_1,b_3) + 2 r_{\pi}
 \Phi_\pi^P(x_3)
E_{ei}(t') h_{e}(x_3,x_1,b_3,b_1) \bigg\}\;,
\label{eq:fepip2}
\en
%%%%%%%%%%
\be
{\cal M}_{e\pi} &=& 32 \pi C_Fm_B^4/\sqrt{2N_C}\int_0^1
dx_1dx_2dx_3 \int_0^{\infty} b_1 db_1\, b_2
db_2\,\Phi_B(x_1,b_1) \Phi_{f}(x_2) \non
&& \cdot \bigg\{ [(1-x_2)\Phi_{\pi}(x_3) -r_{\pi}x_3(\Phi_{\pi}^P(x_3)-\Phi_{\pi}^T(x_3))]
E'_{ei}(t) h_n(x_1,\bar{x}_2,x_3,b_1,b_2)
\nonumber \\
&&\;\;\;- [(x_2+x_3)\Phi_{\pi}(x_3) - r_{\pi} x_3 \left(
\Phi_{\pi}^P(x_3)+\Phi_{\pi}^T(x_3)
       \right]
E'_{ei}(t') h_n(x_i,b_1,b_2) \bigg\}\; ,
\end{eqnarray}
%%%%
\be
{\cal M}^{P1}_{e\pi}&=& \frac{32}{\sqrt{6}} \pi C_Fm_B^4r_{f}\int_0^1
dx_1dx_2dx_3 \int_0^{\infty} b_1 db_1\, b_2 db_2\,\Phi_B(x_1,b_1)
\nonumber \\
& &\cdot \bigg\{ E'_{ei}(t) h_n(x_1,\bar{x}_2,x_3,b_1,b_2)\cdot
\left[ (x_2-1)\Phi_{\pi}^A(x_3) \left( \Phi_{f}^S(x_2) + \Phi_{f}^T(x_2)
\right)
\right.\nonumber \\
& &\;\;\;\;\;\;
+r_{\pi}(x_2-1)\left(
\Phi_{\pi}^P(x_3)-\Phi_{\pi}^T(x_3) \right) \left( \Phi_{f}^S(x_2)+
\Phi_{f}^T(x_2)
\right)\nonumber\\
& &\;\;\;\;\;\;\left. - r_{\pi} x_3 \left(
\Phi_{\pi}^P(x_3)+\Phi_{\pi}^T(x_3) \right) \left( \Phi_{f}^S(x_2)-
\Phi_{f}^T(x_2) \right)\right]
\nonumber \\
& & \;\;\;\;+  E'_{ei}(t') h_n(x_i,b_1,b_2)\cdot \left[
x_2\Phi_{\pi}^A(x_3) \left( \Phi_{f}^S(x_2) - \Phi_{f}^T(x_2)
\right) \right.\nonumber \\
&&\;\;\;\;\; +r_{\pi}x_2\left( \Phi_{\pi}^P(x_3)-\Phi_{\pi}^T(x_3)
\right) \left( \Phi_{f}^S(x_2)- \Phi_{f}^T(x_2)
\right)\nonumber\\
& &\;\;\;\;\;\;\left. + r_{\pi} x_3 \left(
\Phi_{\pi}^P(x_3)+\Phi_{\pi}^T(x_3) \right) \left( \Phi_{f}^S(x_2)+
\Phi_{f}^T(x_2) \right)\right]\bigg\}\;,
\en
\be
{\cal M}^{P2}_{e\pi} &=& -\frac{32}{\sqrt{6}} \pi C_Fm_B^4 \int_0^1
dx_1dx_2dx_3 \int_0^{\infty} b_1 db_1\, b_2
db_2\,\Phi_B(x_1,b_1)\Phi_f(x_2)
\nonumber \\
& &\cdot \bigg\{ \left[ (x_2-x_3-1)\Phi_{\pi}^A(x_3)
+r_{\pi}x_3\left( \Phi_{\pi}^P(x_3)+\Phi_{\pi}^T(x_3) \right)\right]
E'_{ei}(t) h_n(x_1,\bar{x}_2,x_3,b_1,b_2)
\nonumber \\
& & \;\;\;+ \left[x_2\Phi_{\pi}^A(x_3)-r_K
x_3(\Phi_K^P(x_3)-\Phi_K^T(x_3))\right] E'_{ei}(t')
h_n(x_i,b_1,b_2)\bigg\}\;,
\en
%%%
\be
{\cal M}_{a\pi} &=&
\frac{32}{\sqrt{6}}\pi C_Fm_B^4 \int_0^1 dx_1dx_2dx_3 \int_0^{\infty} b_1
db_1\, b_2 db_2\,\Phi_B(x_1,b_1)
\nonumber \\
& &\cdot \left \{ \left[ -x_2\Phi_{\pi}^A(x_3) \Phi_{f}(x_2) \right.\right.
\non
&& \left.\left.
+ r_{\pi}r_{f} \Phi_{f}^T(x_2)
\left( (x_2+x_3-1)\Phi_{\pi}^P(x_3)+(-x_2+x_3+1) \Phi_{\pi}^T(x_3)\right)
\right.\right. \non
& &\left. \left.
+ r_{\pi}r_{f} \Phi_{f}^S(x_2)
\left( (x_2-x_3+3)\Phi_{\pi}^P(x_3)-(x_2+x_3-1) \Phi_{\pi}^T(x_3) \right)\right]
\right.\non
&& \left. \cdot  E'_{ai}(t) h_{na}(x_1,x_2,x_3,b_1,b_2)
\right. \non
& & \left.
-  E'_{ai}(t') h'_{na}(x_1,x_2,x_3,b_1,b_2)\cdot
\left[ (x_3-1)\Phi_{\pi}^A(x_3) \Phi_{f}(x_2)
\right.\right. \non
& &\left. \left.
+r_{\pi}r_{f} \Phi_{f}^S(x_2)
\left( (x_2-x_3+1)\Phi_{\pi}^P(x_3)-(x_2+x_3-1) \Phi_{\pi}^T(x_3) \right)
\right. \right. \non
& &\left. \left. + r_{\pi}r_{f} \Phi_{f}^T(x_2) \left(
(x_2+x_3-1)\Phi_{\pi}^P(x_3)-(1+x_2-x_3) \Phi_{\pi}^T(x_2)
\right)\right] \right \}\;,
\en
\be
 {\cal M}^{P1}_{a\pi} &=& \frac{32}{\sqrt{6}} \pi C_Fm_B^4
 \int_0^1 dx_1dx_2dx_3 \int_0^{\infty} b_1 db_1\, b_2 db_2\,\Phi_B(x_1,b_1)
\nonumber \\
&&\cdot \bigg\{ \left[r_{\pi} (1+x_3)\Phi_{f}(x_2)
(\Phi_{\pi}^T(x_3)-\Phi_{\pi}^P(x_3))
+r_{f}(x_2-2)\Phi_{\pi}(x_3)(\Phi_{f}^S(x_2)+\Phi_{f}^T(x_2))\right]\nonumber \\
&&\;\;\;\;\cdot E'_{ai}(t) h_{na}(x_1,x_2,x_3,b_1,b_2)
\nonumber \\
&& \;\;\;- \left[r_{\pi} (x_3-1)\Phi_{f}(x_2)
(\Phi_{\pi}^T(x_3)-\Phi_{\pi}^P(x_3))+r_{f}x_2\Phi_{\pi}(x_3)(\Phi_{f}^S(x_2)+\Phi_{f}^T(x_2))
\right]\nonumber \\
&&\;\;\;\;\cdot  E'_{ai}(t') h'_{na}(x_1,x_2,x_3,b_1,b_2)
\bigg\}\;,
\en
\be
 {\cal M}^{P2}_{a\pi} &=& -\frac{32}{\sqrt{6}} \pi C_Fm_B^4 \int_0^1
dx_1dx_2dx_3 \int_0^{\infty} b_1 db_1\, b_2 db_2\,\Phi_B(x_1,b_1)
\nonumber \\
&&\cdot \bigg\{ \left[(x_3-1)\Phi_{f}(x_2)
\right.\Phi_{\pi}^A(x_3)+4r_{\pi}r_f\Phi^S_{f}(x_2)\Phi_{\pi}^P(x_3)
+r_{\pi}r_{f}\left((x_2-x_3-1)\left(\Phi_{\pi}^P(x_3)\right.\right.\nonumber
\\&&\left.\;\;\cdot\Phi_{f}^S(x_2)-\Phi_{\pi}^T(x_3)\Phi_{f}^T(x_2)\right.)\left.
-(x_2+x_3-1)(\Phi_{\pi}^P(x_3)\Phi_{f}^T(x_2)-\Phi_{\pi}^T(x_3)\Phi_{f}^S(x_2))\right]\nonumber \\
&&\;\;\;\cdot E'_{ai}(t) h_{na}(x_1,x_2,x_3,b_1,b_2)
\nonumber \\
&& \;\;+ \left[x_2 \Phi_{f}(x_2)
\Phi_{\pi}^A(x_3)-x_2r_{\pi}r_f(\Phi_f^S(x2)+\Phi_f^T(x2))(\Phi_{\pi}^P(x_3)-\Phi_{\pi}^T(x_3))
\right.\nonumber \\ &&\;\;\;\left.
-r_{\pi}r_{f} (1-x_3)(\Phi_f^S(x2)-\Phi_f^T(x2))(\Phi_{\pi}^P(x_3)+\Phi_{\pi}^T(x_3))
\right]\nonumber \\
&&\;\;\;\;\cdot E'_{ai}(t') h'_{na}(x_1,x_2,x_3,b_1,b_2) \bigg\}\;,
\en
\be
F_{a\pi} &=&
-F^{P1}_{a\pi}= 8 \pi C_F m_B^4f_B \int_0^1 dx_2 dx_3
\int_0^{\infty} b_2db_2\, b_3db_3\
\nonumber \\
&& \cdot \bigg\{[(x_3-1)\Phi_{\pi}^A(x_3)\Phi_{f}(x_2) - 2r_{\pi}
r_{f}(x_3-2) \Phi_{\pi}^P(x_3)\Phi_{f}^S(x_2) +2r_{\pi} r_{f}x_3
\Phi_{\pi}^T(x_3)\Phi_{f}^S(x_2) ]
\nonumber \\
&&\;\;\;\cdot E_{ai}(t) h_{a}(x_2,1-x_3, b_2, b_3)
\nonumber\\
&&
 \;\;\;+ [x_2\Phi_{\pi}^A(x_3)\Phi_{f}(x_2)-2r_{\pi}
r_{f}\Phi_{\pi}^P(x_3)((x_2+1)\Phi_{f}^S(x_2)+(x_2-1)\Phi_{f}^T)]\nonumber\\
&& \;\;\;\;\cdot E_{ai}(t') h_{a}(1-x_3,x_2, b_3, b_2)\bigg\}, \en

\be
F^{P2}_{a\pi} &=& -16 \pi C_F m_B^4f_B \int_0^1 dx_2
dx_3 \int_0^{\infty} b_2db_2\, b_3db_3\,  \non
&&
\cdot \bigg\{[r_{\pi}(x_3-1)\Phi_{f}(x_2)(\Phi_{\pi}^P(x_3)+\Phi_{\pi}^T(x_3))
+2r_{f}\Phi_{\pi}(x_3)\Phi_{f}^S(x_2)]E_{ai}(t) h_{a}(x_2,\bar{x}_3, b_2, b_3)\non
&& - [2r_{\pi}\Phi_{\pi}^P(x_3)
\Phi_{f}(x_2)+r_{f}x_2\Phi^A_{K}(x_3)
(\Phi_{f}^T(x_2)-\Phi_{f}^S(x_2))]E_{ai}(t') h_{a}(\bar{x}_3,x_2, b_3,
b_2)\bigg\}\;,
\en
\begin{eqnarray}
F_{ef} &=&F^{P1}_{ef}= 8 \pi C_F m_B^4 f_{\pi}\int_0^1 dx_1 dx_2
\int_0^{\infty} b_1db_1\, b_2db_2\, \Phi_B(x_1,b_1)
\nonumber \\
& & \cdot  \bigg\{ \left[ (1+x_2)\Phi_{f}(x_2)-r_{f}(1-2x_2)
\left( \Phi_{f}^S(x_2)+\Phi_{f}^T(x_2) \right) \right] E_{ei}(t)
h_{e}(x_1,x_2,b_1,b_2)
\nonumber\\
& &\;\;\;\;\;\; -2r_{f} \Phi_{f}^S({x_2})E_{ei}(t')
h_{e}(x_2,x_1,b_2,b_1) \bigg\} \;,\end{eqnarray}
%%%%
%%%%
\begin{eqnarray}
 F^{P2}_{ef}&=&16 \pi C_F
m_B^4 f_{\pi}r_{\pi}\int_0^1 dx_1 dx_2 \int_0^{\infty} b_1db_1\,
b_2db_2\, \Phi_B(x_1,b_1)
\nonumber \\
& &\cdot \bigg\{ -\left[ \Phi_{f}(x_2)+r_{f} \left(
x_2\Phi_{f}^T(x_2)-(x_2+2)\Phi_{f}^S(x_2) \right) \right]
E_{ei}(t) h_{e}(x_1,x_2,b_1,b_2)
\nonumber\\
& &\;\;\;\;\;\; +2r_{f} \Phi_{f}^S({x_2})E_{ei}(t')
h_{e}(x_2,x_1,b_2,b_1) \bigg\} \;,\end{eqnarray}
%%%%
%%%%
\begin{eqnarray}
{\cal M}_{ef} &=& \frac{32}{\sqrt{6}} \pi C_Fm_B^4 \int_0^1 dx_1dx_2dx_3
\int_0^{\infty} b_1 db_1\, b_2 db_2\,\Phi_B(x_1,b_1)
\Phi_{\pi}^A(x_3)
\nonumber \\
&&\cdot
 \bigg\{-[(x_3-1)\Phi_{f}(x_2)-r_{f}x_2(\Phi_{f}^S(x_2)-\Phi_{f}^T(x_2))]
E'_{ei}(t) h_n(x_1,1-x_3,x_2,b_1,b_3)
\nonumber \\
&&+\left[-(x_2+x_3)\Phi_{f}(x_2) -r_{f} x_2
(\Phi_{f}^S(x_2)+\Phi_{f}^T(x_2))
       \right]
E'_{ei}(t') h_n(x_1,x_3,x_2,b_1,b_3) \bigg\}\;,
\end{eqnarray}
\begin{eqnarray}
{\cal M}^{P1}_{ef} &=&
\frac{32}{\sqrt{6}} \pi C_Fm_B^4  r_{\pi}\int_0^1
dx_1dx_2dx_3 \int_0^{\infty} b_1 db_1\, b_3 db_3\,\Phi_B(x_1,b_1)
\nonumber \\
&&\cdot
\bigg\{E'_{ei}(t) h_n(x_1,1-x_3,x_2,b_1,b_3)\cdot
 [(x_3-1)\Phi_f(x_2)(\Phi^P_{\pi}(x_3)+\Phi^T_{\pi}(x_3))\nonumber\\
&&\;\;\;+r_f\Phi_{f}^T(x_2)
((x_2+x_3-1)\Phi^P_{\pi}(x_3)+(-x_2+x_3-1)\Phi_{\pi}^T(x_3))
\nonumber\\
 && \;\;\;+r_{f}\Phi_{f}^S(x_2)((x_2-x_3+1)\Phi_{\pi}^P(x_3)-(x_2+x_3-1)\Phi_{\pi}^T(x_3))]
\nonumber\\
&&\;\;\;+[-x_3\Phi_{f}(x_2)(\Phi^T_{\pi}(x_3)-\Phi_{\pi}^P(x_3))-r_fx_3(\Phi_f^S(x_2)-\Phi_f^T(x_2))
(\Phi_{\pi}^P(x_3)-\Phi_{\pi}^T(x_3))\nonumber\\
&& \;\;\;-r_fx_2
(\Phi_f^S(x_2)+\Phi_f^T(x_2))(\Phi_{\pi}^P(x_3)+\Phi_{\pi}^T(x_3))]
E'_{ei}(t') h_n(x_1,x_3,x_2,b_1,b_3) \bigg\}\;,
\end{eqnarray}
\begin{eqnarray}
{\cal M}^{P2}_{ef} &=& - \frac{32}{\sqrt{6}} \pi C_Fm_B^4 \int_0^1
dx_1dx_2dx_3 \int_0^{\infty} b_1 db_1\, b_2
db_2\,\Phi_B(x_1,b_1)\Phi_{\pi}^A(x_3)
\nonumber \\
& &\cdot \bigg\{ \left[ (x_3-x_2-1)\Phi_f(x_2) -r_{f}x_2\left(
\Phi_f^S(x_2)+\Phi_f^T(x_2) \right)\right] E'_{ei}(t)
h_n(x_1,1-x_2,x_3,b_1,b_2)
\nonumber \\
& & \;\;\;+ \left[x_2\Phi_f(x_2)+r_f
x_2(\Phi_f^S(x_2)-\Phi_f^T(x_2))\right] E'_{ei}(t')
h_n(x_1,x_3,x_2,b_1,b_2)\bigg\}\;,
\end{eqnarray}
\begin{eqnarray}
 {\cal M}_{a} &=& -\frac{32}{\sqrt{6}} \pi C_Fm_B^4 \int_0^1
dx_1dx_2dx_3 \int_0^{\infty} b_1 db_1\, b_3 db_3\,\Phi_B(x_1,b_1)
\nonumber \\
& &\cdot \bigg\{ \left[ x_3\Phi_{\pi}^A(x_3) \Phi_{f}(x_2)
\right.+r_{\pi}r_{f} \Phi_{f}^T(x_2) \left(
(x_2-x_3+1)\Phi_{\pi}^T(x_3)-(x_2+x_3-1) \Phi_{\pi}^P(x_3)
\right)\non
& &\;\;\;\;\;\;\left. + r_{\pi}r_{f} \Phi_{f}^S(x_2) \left(
(-x_2+x_3+3)\Phi_{\pi}^P(x_3)+(x_2+x_3-1) \Phi_{\pi}^T(x_3)
\right)\right]
\non && \qquad \cdot
E'_{ai}(t) h_{na}(x_1,x_3,x_2,b_1,b_3)
\nonumber \\
& & \;\;\;\;\;+  E'_{ai}(t') h'_{na}(x_1,x_3,x_2,b_1,b_3)\left[
(x_2-1)\Phi_{\pi}^A(x_3) \Phi_{f}(x_2)
\right.\nonumber \\
& &\;\;\;\;\;\;+r_{\pi}r_{f} \Phi_{f}^T(x_2) \left(
(-x_2+x_3+1)\Phi_{\pi}^T(x_3)-(x_2+x_3-1) \Phi_{\pi}^P(x_3)
\right)\nonumber\\
& &\;\;\;\;\;\;\left. + r_{\pi}r_{f} \Phi_{f}^S(x_2) \left(
(x_2-x_3-1)\Phi_{\pi}^P(x_3)+(x_2+x_3-1) \Phi_{\pi}^T(x_3)
\right)\right] \bigg\}\;,
\end{eqnarray}
\begin{eqnarray}
{\cal M}^{P1}_{af} &=& \frac{32}{\sqrt{6}} \pi C_Fm_B^4 \int_0^1
dx_1dx_2dx_3 \int_0^{\infty} b_1 db_1\, b_3 db_3\,\Phi_B(x_1,b_1)
\nonumber \\
&&\cdot \bigg\{ \left[r_{f} (x_2+1)\Phi_{\pi}^A(x_3)
(\Phi_{f}^S(x_2)-\Phi_{f}^T(x_2))
+r_{\pi}(x_3-2)\Phi_{f}(x_2)(\Phi_{\pi}^P(x_3)+\Phi_{\pi}^T(x_3))\right]\nonumber \\
&& \;\;\;\;\cdot E'_{ai}(t) h_{na}(x_1,x_3,x_2,b_1,b_3)
\nonumber \\
&& \;\;\;- \left[r_{f} (x_2-1)\Phi_{\pi}^A(x_3)
(\Phi_{f}^S(x_3)-\Phi_{f}^T(x_3))+r_{\pi}x_3\Phi_{f}(x_2)(\Phi_{\pi}^P(x_3)+\Phi_{\pi}^T(x_3))
\right]\nonumber \\
&& \;\;\;\;\cdot E'_{ai}(t') h'_{na}(x_1,x_3,x_2,b_1,b_3)
\bigg\}\;.
\end{eqnarray}

\begin{eqnarray}
F_{af} &=&F^{P1}_{af}= 8 \pi C_F m_B^4f_B \int_0^1 dx_2 dx_3
\int_0^{\infty} b_2db_2\, b_3db_3\
\nonumber \\
&& \cdot  \left \{ \left [ (x_2-1)\Phi_{\pi}^A(x_3)\Phi_{f}(x_2)
+ 2r_{\pi} r_{f}(x_2-2) \Phi_{\pi}^P(x_3)\Phi_{f}^S(x_2) -2r_{\pi} r_{f}x_2
\Phi_{\pi}^P(x_3)\Phi_{f}^T(x_2) \right ]\right.
\non
&&\;\;\;\;\left. \cdot E_{ai}(t) h_{a}(x_3,1-x_2, b_3, b_2)
\right.\non
&& \left.
+ \left [x_3\Phi_{\pi}^A(x_3)\Phi_{f}(x_2)+2r_{\pi}
r_{f}\Phi_{f}^S(x_2)((x_3+1)\Phi_{\pi}^P(x_3)+(x_3-1)\Phi_{\pi}^T(x_3))\right]
\right. \non
&& \,\,\,\left. \cdot E_{ai}(t') h_{a}(1-x_2,x_3, b_2, b_3)\right \},
\end{eqnarray}
\begin{eqnarray}
F^{P2}_{af} &=& 16 \pi C_F m_B^4f_B \int_0^1
dx_2 dx_3 \int_0^{\infty} b_2db_2\, b_3db_3\,  \nonumber\\
&&
\cdot \left \{ \left [
r_{f}(x_2-1)\Phi_{\pi}^A(x_3)(\Phi_{f}^S(x_2)+\Phi_{f}^T(x_2))
-2r_{\pi}\Phi_{\pi}^P(x_3)\Phi_{f}(x_2)\right ]\right.\non
&&\left.
 \quad \cdot E_{ai}(t) h_{a}(x_3,\bar{x}_2, b_2, b_3) \right. \non
&& \;\;\;\left.
- \left [2r_{f}\Phi^A_{K}(x_3) \Phi_{f}^S(x_2)+r_{\pi}x_3\Phi_{f}(x_2)
(\Phi_{\pi}^P(x_3)-\Phi_{\pi}^T(x_3)) \right ]
\right.\non
&&\left. \quad \cdot
E_{ai}(t') h_{a}(1-x_2,x_3, b_2, b_3)\right \},
\label{eq:fafp2}
\end{eqnarray}
where $r_f=m_f/m_B$ and $r_{\pi}=m^{\pi}_0/m_B$. The explicit expressions of
hard functions $E_{ei,ai}^{(\prime)}(t)$ and $h_{e,a}(x_i,b_j), \cdots$
can be found for example in Ref.\cite{guodq07}.
For $\bar{B}^0 \to f_0(980)\etapp$ decays, one can find the corresponding decay amplitudes
from those given in Eq.~(\ref{eq:fepip2}-\ref{eq:fafp2}) by simple replacements.

\section*{2. Numerical results and discussions}

For numerical calculation, we will use the following input parameters:
\be
m(f_0(980))&=&0.98 {\rm GeV}, \quad  m_\pi=0.14 {\rm GeV},
\quad m_{\eta}=547.5{\rm MeV},\quad  m_{\eta^{\prime}}=957.8{\rm MeV},\non
\quad M_B &=& 5.28 {\rm GeV}, \quad
 m_0^\pi = 1.4 {\rm GeV},\quad    M_W = 80.42 {\rm GeV}, \quad
 \bar{f}_{f_0}=(0.37 \pm 0.02){\rm GeV} \non
f_B &=& 0.19 {\rm  GeV},  \quad  f_{\pi} = 0.13  {\rm GeV},
\quad \tau_{B^\pm}=1.671\; ps, \quad \tau_{B^0}=1.536 \; ps,\non
V_{tb}&=&0.9997, \quad |V_{td}|=0.0082, \quad V_{ud}=0.974, \quad
|V_{ub}|=0.00367,
\en
with the CKM angle $\beta=21.6^\circ$ and $\gamma=60^\circ$.

\begin{table}
\caption{The pQCD predictions (in unit of $10^{-6}$) for the branching ratios  of
$B\to f_0(980)\pi, f_0(980)\etapp$ decays.}
\label{branchs}
\begin{center}
\begin{tabular}{c|c|c|c|c}   \hline \hline
   Channel & $\theta_{1}=32.5^\circ \pm 7.5^\circ $ & $\theta_{2}=152.5^\circ \pm 12.5^\circ$ &
   Data\cite{hfag} &QCDF\cite{hycheng1}  \\   \hline
$Br(B^-\to f_0(980) \pi^-)$      &$2.5\pm 1.0 $ & $1.6^{+1.8}_{-0.6}$    & $<3.0$ &$0.9$\\
$Br(\bar B^0\to f_0(980)\pi^0) $ &$0.26 \pm 0.06$ & $0.04^{+0.06}_{-0.02}$ & --     &$0.03$\\
$Br(\bar B^0\to f_0(980)\eta) $  &$0.25 \pm 0.07$ & $0.59\pm 0.20$ & $<0.4$ &--\\
$Br(\bar B^0\to f_0(980)\etap)$  &$0.67 \pm 0.06$ & $0.26 \pm 0.03$ & $<1.5$ &--\\
\hline\hline
\end{tabular}  \end{center}
\end{table}

It is straightforward to calculate the branching ratios of the
considered decays.
If $f_0(980)$ is purely composed of $\bar nn$, the pQCD predictions for the branching ratios
are
\be
{\cal B}(\bar B^0\to f_0(980)\pi^0)&=&(0.89^{+0.10+0.16+0.05}_{-0.08-0.13-0.03})\times 10^{-6},\non
{\cal B}( B^-\to f_0(980)
\pi^-)&=&(16.4^{+1.7+1.1+0.8}_{-1.6-1.2-0.9})\times 10^{-6},\non
{\cal B}(\bar B^0\to
f_0(980)\eta)&=&(2.0^{+0.2+0.4+0.1}_{-0.2-0.3-0.1})\times 10^{-6},\non
{\cal B}(\bar B^0\to f_0(980)
\etap)&=&(1.3^{+0.2+0.3+0.0}_{-0.1-0.2-0.1})\times 10^{-6},
\label{eq:br-nn}
\en
where the theoretical uncertainties are from the decay constant of
$\bar{f}_{f_0}=0.37\pm 0.02$ {\rm GeV},
the Gegenbauer moments $B_1=-0.78 \pm 0.08$ and $B_3=0.02\pm 0.07$.
If $f_0(980)$ is purely composed of $\bar ss$, the branching ratios will be
\be
{\cal B}(\bar B^0\to
f_0(980) \pi^0)&=&(4.66^{+0.52+1.01+0.10}_{-0.49-0.90-0.06})\times 10^{-8},\non
{\cal B}( B^-\to f_0(980) \pi^-)&=&(8.56^{+1.80+2.77+0.96}_{-0.21-1.04-0.00})\times 10^{-8},\non
{\cal B}(\bar B^0\to
f_0(980)\eta)&=&(0.24^{+0.02+0.02+0.05}_{-0.03-0.03-0.03})\times 10^{-6},\non
{\cal B}(\bar B^0\to f_0(980)
\etap)&=&(0.38^{+0.05+0.04+0.04}_{-0.04-0.03-0.03})\times 10^{-6},
\label{eq:br-ss}
\en
where the theoretical uncertainties are from the same hadron parameters as above.

When $f_0(980)$ is treated as a mixing state of $\bar nn$ and $\bar ss$,
the leading order pQCD predictions are listed in Table I, where the two ranges of the
mixing angle $\theta$, $\theta_1=[25^\circ, 40^\circ]$ and $\theta_2=[140^\circ,
165^\circ]$, are taken into account. The QCDF predictions as given in Ref.\cite{hycheng1}
are also listed in Table I as a comparison.
The remaining theoretical uncertainties induced by the errors of other input
parameters and the wave functions are generally $30-50\%$, and not
shown here explicitly.

\begin{figure}[tb]
\begin{center}
\includegraphics[scale=0.72]{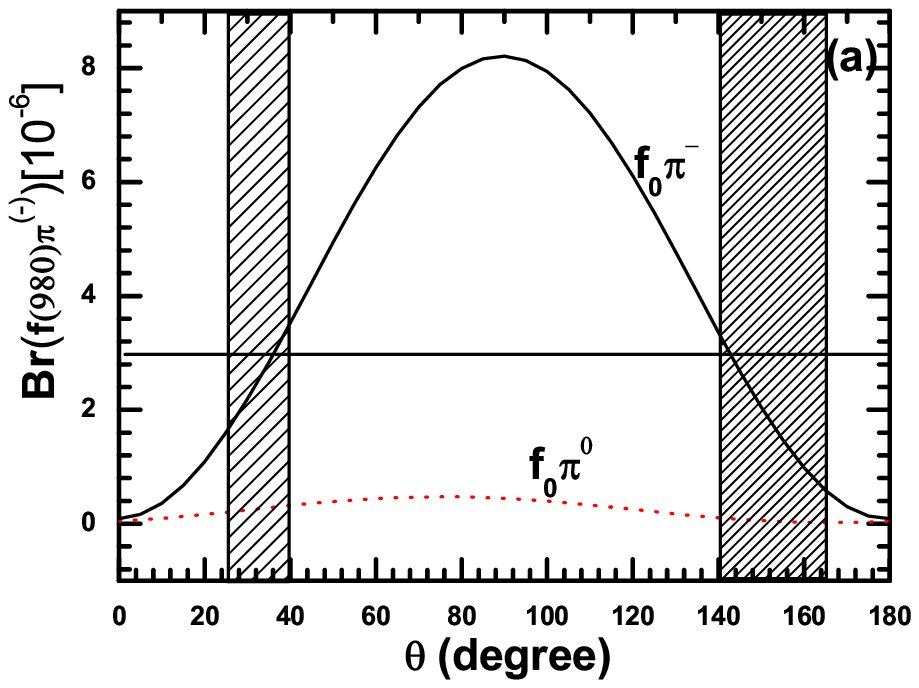}
\includegraphics[scale=0.72]{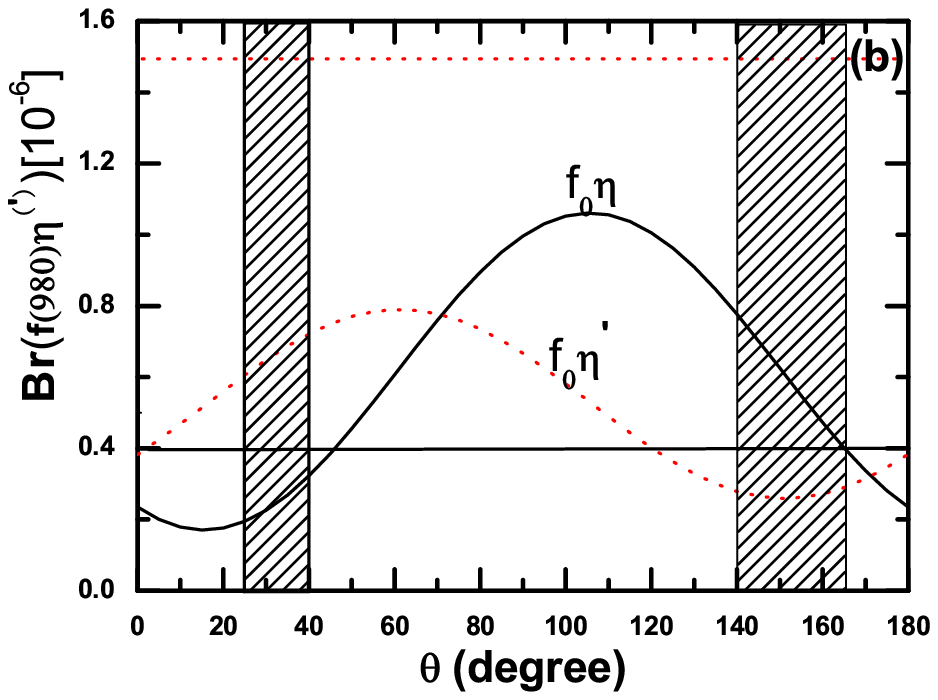}
\vspace{-0.3cm}
\caption{The $\theta-$dependence of the central values of the
pQCD predictions for the branching ratios of (a) $B\to f_0(980) \pi$ decays, and  (b)
$\bar B^0\to f_0 \etapp$ decays.}
\label{f0pi}
\end{center}
\end{figure}

In Fig.~\ref{f0pi}, we show the $\theta-$dependence of the central values of the
pQCD predictions for the branching ratios of the four considered decays.
One should note that the large theoretical uncertainties of the pQCD predictions are not
shown here explicitly.
The two vertical bands show the two ranges of the mixing angle
$\theta$ preferred by the known experiments \cite{hycheng2}, while
the three horizontal solid or dots lines show the corresponding
experimental upper limits \cite{hfag} as listed in Table I. From the
numerical results as shown in Table I and Fig.\ref{f0pi}, one can
not distinguish two regions of the mixing angle $\theta$
from currently available data,
if the still large theoretical uncertainties are taken into account.

%=================================================================

Now we turn to the evaluations of the CP-violating asymmetries of
$B\to f_0(980)\pi,f_0(980)\etapp$ decays in the pQCD approach.
The pQCD predictions for the direct CP-violating asymmetries of the four considered
decays are listed in Table II.
Although the CP-violating asymmetries are large in size, it is still
difficult to measure them, since their branching ratios are generally very
small, say around $10^{-6} \sim 10^{-8}$.

\begin{table}
\caption{The pQCD predictions (in units of $10^{-2}$) for the CP-violating asymmetries of
$B\to f_0(980)\pi, f_0(980)\etapp$ decays.}
\label{acp}
\begin{center}
\begin{tabular}{c|c|c|c|c} \hline \hline
 &  \multicolumn{2}{|c}{$A_{CP}^{dir}$} & \multicolumn{2}{|c}{$A_{CP}^{mix}$} \\
 \cline{2-5}
Channel & $\theta_{1}=[25^\circ,40^\circ]$ & $\theta_{2}=[140^\circ,165^\circ]$
& $\theta_{1}=[25^\circ,40^\circ]$ & $\theta_{2}=[140^\circ,165^\circ]$  \\   \hline
$     B^-\to f_0(980) \pi^-$ &$[50, 64]    $ & $[-39,  7.0]$ & $--$          &$--$\\
$\bar B^0\to f_0(980)\pi^0 $ &$[-7.5, -2.3]$ & $[-99,  -56]$ & $\sim -69 $   &$[-25, 7.1]$\\
$\bar B^0\to f_0(980)\eta  $ &$[-43, -5.0] $ & $[-55,  -30]$ & $[-72, 12]$   &$[-63, -23]$\\
$\bar B^0\to f_0(980)\etap $ &$[-42, -28]  $ & $[-29,  8.5]$ & $[-57,-38]$   &$[-75, -38]$\\
\hline\hline
\end{tabular}  \end{center}
\end{table}

In this paper, based on the assumption of two-quark structure of the scalar meson
$f_0(980)$, we calculated the branching ratios and CP-violating
asymmetries of the four $ B\to f_0(980)\pi$ and $\bar B^0\to f_0(980)\etapp$ decays
by employing the leading order pQCD factorization approach.
The pQCD predictions are generally consistent with both the QCDF predictions and
the currently available experimental upper limits.

\section*{Acknowledgment}
This work is partly supported  by the National Natural Science
Foundation of China under Grant No.10575052 and 10735080.

%%%%%%%%%%%%%%%%%%%%%%%%%%%%%%%%%%%%%%%%%%%%%%%%%%%%%%%%%%%%%%%%%%%%%%%%
%                               references
%%%%%%%%%%%%%%%%%%%%%%%%%%%%%%%%%%%%%%%%%%%%%%%%%%%%%%%%%%%%%%%%%%%%%%%%

\end{document}